\documentclass[preprint,preprintnumbers,amsmath,amssymb]{revtex4}

\usepackage{amsmath,amssymb}
\usepackage{graphicx}
\usepackage{epsfig}
\usepackage{dcolumn}
\usepackage{bm}

\begin{document}
\title{Circuit architecture explains functional similarity of bacterial heat shock responses} 
\author{Masayo Inoue}
\affiliation{Cybermedia Center, Osaka University, Toyonaka, Osaka 560-0043, Japan}
\affiliation{
Center for Models of Life, Niels Bohr Institute, University of Copenhagen, Blegdamsvej 17, Copenhagen, 2100-DK, Denmark. 
}
\author{Namiko Mitarai and Ala Trusina}
\affiliation{
Center for Models of Life, Niels Bohr Institute, University of Copenhagen, Blegdamsvej 17, Copenhagen, 2100-DK, Denmark. 
}

\begin{abstract}
Heat shock response is a stress response to temperature changes and a 
consecutive increase in amounts of 
unfolded proteins. To restore homeostasis, cells upregulate 
chaperones facilitating protein folding by means of transcription factors (TF). 
We here investigate two heat shock systems: one characteristic to gram negative bacteria, mediated by transcriptional 
activator $\sigma^{32}$ in \textit{E. coli}, and another characteristic to gram 
positive bacteria, mediated by transcriptional repressor HrcA in \textit{L. lactis}. 
We construct simple mathematical model of the two systems focusing on 
the negative feedbacks, where free chaperons suppress $\sigma^{32}$ 
activation in the former, while they activate HrcA repression in the latter.  
We demonstrate that both systems, in spite of the difference at the TF regulation level, 
are capable of showing very similar heat shock dynamics. 
We find that differences in regulation impose distinct constrains on chaperone-TF binding affinities: 
the binding constant of free $\sigma^{32}$ to chaperon DnaK, known to be in 
$100$ nM range, set the lower limit 
of amount of free chaperon that the system can sense the change at the heat shock,
while the binding affinity of HrcA to chaperon GroE set the upper 
limit and  have to be rather large extending into the micromolar range.
\end{abstract}
\maketitle

\section*{Introduction}

Cellular homeostasis is essential for proper protein folding and function.
The perturbations to homeostasis, e.g. due to change in temperature or osmotic pressure, result in protein unfolding or/and misfolding. Heat shock, i.e., sudden increase of temperature, causes such protein unfolding and misfolding and can result in cell death.
To counteract the heat shock, cells upregulate production of chaperons and proteases -- enzymes that help folding the unfolded proteins and degrade terminally misfolded proteins, respectively.
The heat shock response is one of the stress responses characteristic for nearly all living organisms.
Interestingly, the protein sequence of most chaperones and proteases is well conserved from bacteria to humans \cite{Kueltz2003}.
It is, however, unclear if the features of heat shock response are also preserved at the level of the  architecture of regulatory circuits governing heat shock response.

In this article we attempt to answer this question and derive useful insights by comparing the heat shock in \textit{E. coli} and \textit{L. lactis}  These organisms utilize two different mechanisms;  
a system with $\sigma^{32}$ and DnaK in \textit{E. coli }  and a system with HrcA and GroE in \textit{L. lactis}. Both mechanisms are widely observed in microorganisms. 
A transcriptional activator RpoH, $\sigma^{32}$ homolog, is found in the alpha-, beta-, and gamma-proteobacteria, while a transcriptional inhibitor, HrcA, is widely distributed in eubacteria but not in the gamma-proteobacteria. 
Interestingly, there also exist bacteria which have both systems and, furthermore, a regulatory loop between $\sigma^{32}$ and HrcA is predicted in some beta-proteobacteria \cite{Permina2003}.

\begin{figure}[t!]
\centerline{\epsfig{file=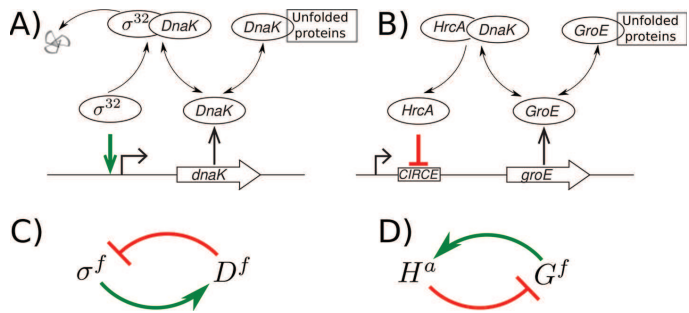,angle=0,width=9cm}}
 \caption{Comparison of the reaction mechanisms between (A,C) $\sigma^{32}$-DnaK system and (B,D) HrcA-GroE system. Views for gene regulations and protein reactions (A and B) and outline illustrations for distinct reactions (C and D). }
  \label{fig:reactcompare}
\end{figure}

Heat shock responses have been extensively studied both experimentally and theoretically.
In experiments, protein sequences and their regulatory mechanisms are revealed in the both systems (figure~\ref{fig:reactcompare}) \cite{Narberhaus1999}.
While $\sigma^{32}$ system is modeled to quite large extent \cite{El-Samad2005, Arnvig2000}, to our knowledge there is no modeling work on HrcA system.
Our aim in this study is to construct a simple model based on known experimental data for each system and theoretically investigate similarities and differences in the regulatory features and the dynamical responses mediated by $\sigma^{32}$ and HrcA.

 One of the striking similarities is emerging at the level of the dynamics of the transcription regulators: $\sigma^{32}$ and HrcA. Both systems respond with a sharp peak in the rate of production of new chaperones: upon a temperature shift, a fast increase up to 4-5 fold within  5-10 minutes (corresponding to about 0.1 generation time \cite{Arnvig2000}) is followed by a rapid decline to a new steady state that is about $1.5$ fold of the one at the starting temperature in both $\sigma^{32}$ and HrcA system \cite{Arnvig2000, Wetzstein1992}.
 This similarity is particularly interesting as the mechanisms of transcriptional regulation are very different: while $\sigma^{32}$
is an activator, HrcA is a repressor (see figure~\ref{fig:reactcompare}). 

\section*{Model}

In this section, we explain how we construct our models based on existing experimental observations.

\subsubsection*{$\sigma^{32}$-DnaK system}

Our model with $\sigma^{32}$ and DnaK is in large similar to that outlined in \cite{El-Samad2005} with exception of the differences outlined below.
The main players in the model are: $\sigma^{32}$ ($\sigma$; transcription factor), DnaK ($D$; chaperon), and unfolded proteins ($U$).
(In the following we will denote total concentration of protein $X$ as $[X^t]$ and the concentration of free protein as $[X^f]$.)

$\sigma^{32}$ is unstable and is present only in small amounts ($[\sigma^t] \sim 200$ proteins per cell).
Under steady state conditions it is sequestered by chaperones, such that the most $\sigma^{32}$ exists as a complex ($[\sigma ^t] \sim [\sigma D]$).  For simplicity, we assume it is produced at a constant rate independent of temperature.  
In addition, for a simpler comparison with HrcA system, we do not include the stabilization of $\sigma^{32}$ (half-life changes from $1$ to $8$ min) during the heat shock. These two are the main differences from the model in \cite{El-Samad2005}.

When not bound to DnaK, $\sigma^{32}$ forms a complex with RNA polymerase (RNAp) and targets RNAp for transcription of heat shock proteins, including DnaK.  For simplicity we will  refer to  this $[RNAp:\sigma^{32}]$ complex as "free" $\sigma^{32}$, $[\sigma^f]$ (i.e. not bound to DnaK).

Being a chaperone, DnaK facilitates proteins folding and thus forms transient complexes with unfolded proteins.
A temperature shift destabilizes the folding of existing folded proteins and also hinders folding of de novo synthesized proteins, thus sequestering all the existing chaperones and creating the demand for additional chaperones. The demand for additional chaperones is sensed and mediated by $\sigma^{32}$: as long as there are enough unfolded proteins to keep chaperones sequestered away from $\sigma^{32}$, it will facilitate transcription of heat shock proteins.
Namely, the regulatory network has a negative feedback loop (see figure~\ref{fig:reactcompare}A and C), i.e., $\sigma^{32}$ activates DnaK by transcriptional activation ("slow" reaction) while DnaK inhibits $\sigma^{32}$ by complex formation ("fast" reaction).

We assume the reactions $\sigma ^f + D^f \rightleftharpoons \sigma D$ and $U^f + D^f \rightleftharpoons UD$ are in equilibrium as the kinetics of complex formation and dissociation between $\sigma^{32}$ and DnaK and between unfolded proteins and DnaK are much faster than transcription and translation.
Based on these observations and assumptions, 
we describe the system's dynamics with the following equations;
\begin{align}
    \dot{[\sigma^t]} &= \eta -\gamma_s [\sigma^t] -\gamma_c [\sigma D], \label{s1} \\
    \dot{[D^t]} &= \alpha_d \frac{ [\sigma^f]/K_{\sigma}}{1+ [\sigma^f]/K_{\sigma}} -\gamma_s [D^t], \label{s2} \\
    \dot{[U^t]} &= F(T) -\gamma_{us} [UD], \label{s3} \\
    [\sigma D] &=\frac{[\sigma^f][D^f]}{K_j}, \label{s4}\\
    [UD] &=\frac{[U^f] [D^f]}{K_k}, \label{s5}
\end{align}
with the conditions of mass conservation
\begin{align}
     [\sigma ^t] &= [\sigma ^f]+[\sigma D] \sim 200 \mbox{ nM}, \label{s6} \\
     [D^t] &= [D^f]+[\sigma D]+[UD] \sim 20,000 \mbox{ nM}, \\
     [U^t] &= [U^f]+[UD].
\end{align}

 The rate of change in $\sigma^{32}$ (eq.~(\ref{s1})) is given by  constitutive transcription rate, $\eta$, dilution due to cell division with the cell doubling, $1/\gamma_s$, and degradation with fast rate, $\gamma_c$. Since $\sigma^{32}$ is degraded mainly through chaperone-dependent degradation by FisH \cite{Gamer1996},  the fast degradation term depends on the complex $[\sigma D]$.
The production of chaperon DnaK (first term in eq.~(\ref{s2}))  transcriptionally activated by free $\sigma^{32}$, $[ \sigma^f]$, and parameterized by  the maximal production rate  $\alpha_d$. For $[ \sigma^f]$ dependence, we adopt the Michaelis-Menten form with dissociation constant $K_\sigma$. 

Chaperones are typically stable proteins with the half life comparable to cell doubling time $1/\gamma_s$, which is included in the second term of   eq.~(\ref{s2}).
The first term in eq.~(\ref{s3}), $F(T)$, represents that  the production of unfolded protein is increasing function of the temperature, $T$.  Unfolded proteins can re-fold correctly with help of chaperons and are thus removed at a rate proportional to $[UD]$ with the rate $\gamma_{us}$ (the second term in eq.~(\ref{s3})).
Eqs.~(\ref{s4}) and (\ref{s5}) represents complex formations in equilibrium, with $K_j$ and $K_k$ are the dissociation constants between free $\sigma^{32}$ and free DnaK  and  between free unfolded protein and free DnaK, respectively \footnote{ In the numerical simulation of the model, the complex formation were solved by using ordinary differential equations with much faster time scales than eqs.~(\ref{s1}) to (\ref{s3}) for the simplicity of calculation.
}. 

 We fix some of the parameter  values according to the experimental observations as follows.
We set $\gamma_c$ to a unit time ($=1$), which shows fast degradation of a complex $\sigma D$ as most $\sigma^{32}$ exist as the complex in the steady state.  The time scale of the fast degradation, 1/$\gamma_c$ is assumed to be around $1$ min \cite{Straus1987} \footnote{From eqs.~(\ref{s1}) and (\ref{s4}), the degradation term is given as $\gamma_c [\sigma_t]/(1+K_j/[D_f])$. Noting $K_j/[D_f] \sim 1/10$ (see Table~\ref{tbl:parameter} and figure~\ref{fig:function}), $1/\gamma_c$ approximately gives the lifetime of $\sigma^{32}$.}.
$1/\gamma_s$ is a time scale for slow degradation of $\sigma ^t$ and $D^t$ and it is set to the inverse number of cell division time ($\sim 30$).
 We estimate $\eta\sim 200$ nM/min from an observation that $[\sigma ^t] \sim 200$ nM. 
The dissociation constant between chaperones and $\sigma^{32}$ range between $5$ $\mu$M and $19$ nM between correspondingly DnaK/DnaJ and $\sigma^{32}$ \cite{Gamer1996}.
We have set the constant to be $K_j =100$ nM, following the choice in \cite{El-Samad2005}. 

The rest of parameters, $\alpha_d, K_{\sigma}, K_k, \gamma_{us}$, and $F(T)$, have been chosen so that the model reproduces experimental observations, i.e. (i)
$[D^t] \sim 20,000$ nM \cite{Pedersen1978, Maaloe1966}.
(ii) DnaK production rate 
changes $4 \sim 6$ times for a peak and its new steady state after heat shock becomes $1.5$ times of the before heat shock \cite{Arnvig2000}.
(iii) The peak time is less than $5$ time units 
\footnote{In the original experiments by Arnvig et al., the response to temperature shift from $30$ to $37 ^{\circ}C$ was peaking at about $0.1$ generations, corresponding to $3-4$ minutes. To address other temperature shifts, the time in our simulations can be rescaled in terms of cell generations such that the key criteria and the main results would still hold. } and the peak shape of the DnaK production rate is symmetric \cite{Straus1987,Arnvig2000, Wetzstein1992}.
(iv) The steady state amount of free unfolded proteins should be kept small both before and after heat shock.
This is affected by $K_k, \gamma_{us}$, and $F(T)$, the properties related to unfolded proteins. 
In this paper we fix $K_k=1$ nM so that $U_f$ is in nanomolar range in steady state, and fit the other two parameters.  We tested higher values of $K_k$ (up to $K_k=1,000$ nM, which would bring $U_f$ to micromolar range) and they all give a proper response as long as $F(T)$ and $\gamma_{us}$ adjusted properly to account for the timing of the peak.

\subsubsection*{HrcA-GroE system}

Next, we construct a model for HrcA and GroE system, where GroE ($G$) is a chaperon and HrcA ($H$) transcriptionally represses GroE.
We adopt the reaction mechanism shown in figure 2 in \cite{Schumann2003}; HrcA repressor is released from the ribosomes as an inactive protein ($H^i$), which can not bind to its operator, and it has to interact with the GroE chaperonin system to become active ($H^a$).
The inactive HrcA ($H^i$) interacts with chaperon GroE ($HG$) and the active HrcA is released.
The active HrcA ($H^a$) is able to bind to its operators and transcriptionally inhibits the production of GroE, while the active HrcA becomes inactive again at a constant rate, i.e., upon dissociation from its binding site, HrcA is in its inactive form again \cite{Lopez2012}.
In this model, we assume that the total amount of HrcA ($[H^t]=[H^a]+[H^i]+[HG]$) is a constant for simplicity.
As in case with DnaK, GroE chaperons makes a complex with an unfolded protein and helps it to refold correctly.
Similar to $\sigma^{32}$ system, we assume that the two reactions ($H^i+ G^f \rightleftharpoons HG$ and $U^f + G^f \rightleftharpoons UG$) are fast compared with other reactions and always in the equilibrium states.

This system also has a negative feedback loop between transcription factor HrcA and chaperon GroE. However, the regulation is opposite; the active HrcA inhibits GroE with a slow reaction (transcriptional inhibition) and GroE activates the inactive HrcA with a fast reaction (enzymatic modification).
From the points described, we obtain the following reaction equations; 
\begin{align}
    \dot{[H^a]} &= \beta_h [HG] - \gamma_c [H^a], \label{h1} \\
    \dot{[G^t]} &= \beta_g\frac{1}{1+ [H^a]/K_h} -\gamma_s [G^t], \label{h2} \\
    \dot{[U^t]} &= F(T) -\gamma_{uh} [UG], \label{h3} \\
   [HG] &= \frac{[H^i][G^f]}{K_l}, \label{h4} \\
   [UG] &= \frac{[U^f][G^f]}{K_m}, \label{h5}
\end{align}
with the conditions of mass conservation
\begin{align}
     [H^t] &= [H^a]+[H^i]+[HG], \label{h6}\\
     [G^t] &= [G^f]+[HG]+[UG], \\
     [U^t] &= [U^f]+[UG].
\end{align}

 Eq.~(\ref{h1}), represents the time evolution of $[H^a]$ with constant total amount of HrcA,
where the production rate of $[H^a]$ is given by activation from $H^i$ through forming a complex $[HG]$ with the rate $\beta_h$, and inactivation happens with a fast rate $\gamma_c$.
The time evolutions of chaperon $[G^t]$ (eq.~(\ref{h2})) and of unfolded protein $[U^t]$ (eq.~(\ref{h3})) are similar to eqs.~(\ref{s2}) and (\ref{s3}), except that $[G^t]$ is transcriptionally inhibited by $[H^a]$ in eq.~(\ref{h2}). Here, $\beta_g$ is the maximum production rate of GroE, 
$K_h$ is the dissociation constant of active HrcA to GroE promoter, and 
$\gamma_{uh}$ is the rate of removal of unfolded proteins by GroE. 
Eqs.(\ref{h4}) and (\ref{h5}) represents complex formations in equilibrium, with $K_l$ and $K_m$ are the dissociation constants between inactive HrcA and free GroE and between free unfolded protein and free GroE, respectively

Most of the parameters were not experimentally measured for the HrcA-GroE system.
However, for a fair comparison between the two systems, we, whenever possible, use same parameter values as in $\sigma^{32}$-DnaK system (see Table~\ref{tbl:parameter} for correspondence relation).
Thus we assume similar concentrations for transcription factor $[H^t] = 200$ nM ($\sim [\sigma ^t]$) and chaperones $[G^t] \sim 20,000$ nM ($\sim [D^t]$).
Although these numbers have not been validated experimentally, they match the typical concentrations of transcription factors (of order $100$ nM) and chaperones (of order $10$ $\mu$M).

The fastest doubling time in \textit{L. lactis} is also about $30$ min \cite{Holubova2007}, thus we use the same doubling time as in \textit{E. coli} with a corresponding doubling time $\gamma_s=0.03$.
Unlike $\sigma^{32}$, HrcA is a stable protein (in \textit{B. subtilis}), with the half-life more than $60$ minutes \cite{Wiegert2001}. However, it has been suggested that HrcA is present in two conformations, one is active and another is inactive,  and the equilibrium between these two states is modulated by GroEL, which shift the equilibrium towards the active state \cite{Lopez2012}.
 Thus $\gamma_c$, which we set to be $1$, is representing the rate of conversion from active to inactive state, rather than protein half-life as is the case in $\sigma_{32}$.

The time of the peak, fold induction at the peak and fold change of the new steady state are overall similar in the activity of CIRCA operon \cite{Wetzstein1992} (corresponding to GroEL production rate) and $\sigma^{32}$ governed chaperone (DnaK) production. This allows us to use the same criteria (i) - (iv)  to set
$K_m$, $\beta_g, K_h, \gamma_{uh}, \beta_h, F(T)$, and $K_l$.
We study the response of this model assuming the same conditions as for $\sigma^{32}$ system (figure~\ref{fig:direct} C,D).
(In HrcA model, we assumed $[H^t]$ is constant for simplicity. However, the model also works even if we discard this postulation and include that  HrcA inhibits its own transcription \cite{Schumann2003}, i.e., the time evolution of $[H^t]$ will be described as $\dot{[H^t]} =\mu /(1+ [H^a]/K_t) -\gamma_s [H^t]$.)

\begin{table}[h]
\begin{tabular}{|c c|c||c c|c|} \hline
\multicolumn{3}{|c||}{$\sigma^{32}$-DnaK system} & \multicolumn{3}{|c|}{HrcA-GroE system}  \\ \hline
$\eta$                  &   & 200   & $[H^t]$                 &   & 200       \\ \hline
$\gamma_c$       &   & 1       &        &   & 1           \\ \hline
$\gamma_s$       &   & 0.03  &         &   & 0.03      \\ \hline
$K_k$                 & $\ast$ & 1        & $K_m$                 & $\ast$ & 1           \\ \hline
$K_j$                  &   & 100    & $K_l$                   & $\ast$ & 100,000 \\ \hline
$\alpha_d$         & $\ast$ & 2,000 & $\beta_g$            & $\ast$ & 2,000     \\ \hline
$K_s$                 & $\ast$ & 10      & $K_h$                 & $\ast$ & 10          \\ \hline
$\gamma_{us}$  & $\ast$ & 0.5     & $\gamma_{uh}$  & $\ast$ & 0.5         \\ \hline
$\ $                     &   & $\ $     & $\beta_h$           & $\ast$ & 3            \\ \hline
$F(T)$ & $\ast$ & 3000 $\rightarrow$ 9000 &  & $\ast$ & 3000 $\rightarrow$ 9000 \\ \hline
\end{tabular}
\caption{Parameter values used in the model. $\ast$ marks parameters chosen such as to reproduce rapid transient response measured in \cite{Arnvig2000,Wetzstein1992}.The rest of the parameters are based on experimental data as described in the text.}
\label{tbl:parameter}
\end{table}

\section*{Results}

\subsection*{Model predicts much weaker binding affinity between HrcA and chaperones.}

\begin{figure}[h]
\begin{center}
\centerline{\epsfig{file=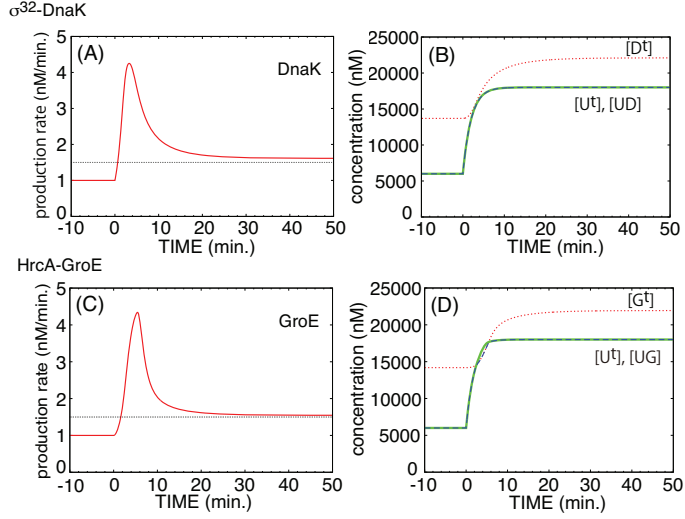,angle=0,width=9cm}}
\caption{Heat shock response in $\sigma^{32}$-DnaK system (A,B) and HrcA-GroE system (C,D). Heat shock is induced at $t=0$. (A,C) shows the  time evolution of the chaperone production rate normalized by the pre-stimulus level.  (B,D) shows the time evolution of the density of total chaperones(dotted lines), total unfolded-protein(bold lines), and complex formed by chaperon and unfolded proteins(broken lines). Parameters used in simulations are shown in Table~\ref{tbl:parameter}.}
  \label{fig:direct}
\end{center}
\end{figure}

In figure~\ref{fig:direct} we show that both $\sigma^{32}$-DnaK and HrcA-GroE systems are able to reproduce experimental observations. Production of chaperones shows characteristic sharp peak with a fast increase up to $4-6$ fold within $5$ minutes and a following decline to a new steady state that is about $1.5$ fold of the pre-stimulus one.
When choosing unknown parameters, our initial strategy was to use the same values for corresponding parameters in each of the systems (see Table~\ref{tbl:parameter}).
Remarkably, this was possible for all but one parameter: the binding affinity of TF to chaperons, $K_j$ and $K_l$.
It appears that while $\sigma^{32}$ binds tightly to the chaperons ($K_j
=100$ nM), it is essential that HrcA is bound only weakly with a micromolar binding constant ($K_l=100$ $\mu$M).

We next study a response upon an inverse heat shock, i.e., a response when temperature is suddenly decreased.
Inverse heat shock response has been studied experimentally in \textit{E. coli} \cite{Arnvig2000, Guisbert2008}(to our knowledge no data exist for \textit{L. lactis}) and is characterized by a rapid decrease in chaperone production with a consequent slow increase to a new steady state that is lower than before temperature decrease.

\begin{figure}[h]
\begin{center}
\centerline{\epsfig{file=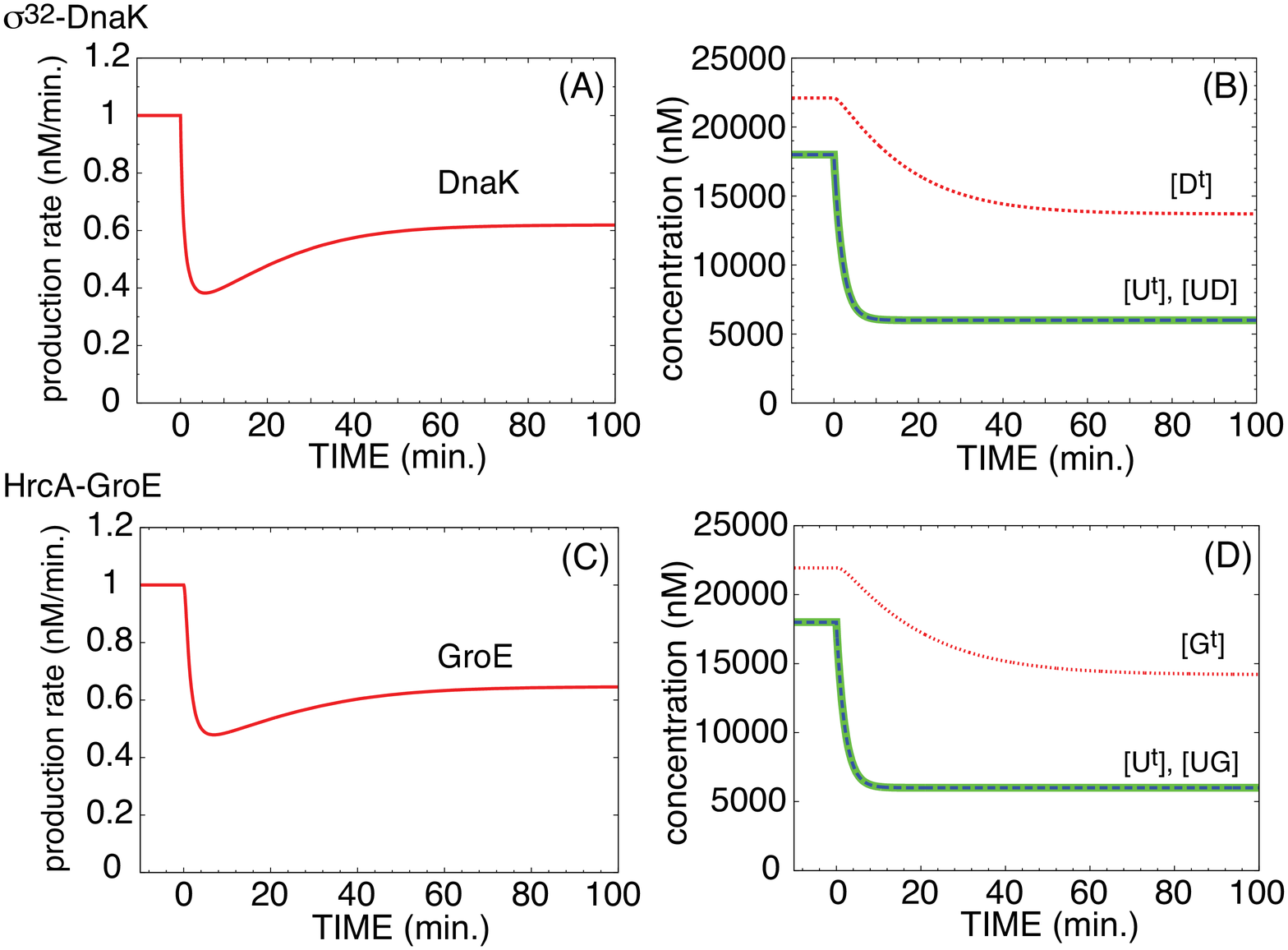,angle=0,width=9cm}}
\caption{Inverse heat shock response in $\sigma^{32}$-DnaK system (A,B) and HrcA-GroE system (C,D). (A,C) shows the time evolution of chaperone production rate normalized by the pre-stimulus level. (B,D) shows the time evolution of the density of total chaperones(dotted lines), total unfolded-protein(bold lines), and complex formed by chaperon and unfolded proteins(broken lines). Parameters used in simulations are shown in Table~\ref{tbl:parameter} except for production rate of unfolded proteins, $F(T)$, which decreases at $t=0$, $F(T)=9,000 \rightarrow 3,000$. }
  \label{fig:SHinverse}
\end{center}
\end{figure}

 In figure~\ref{fig:SHinverse}, we show the chaperon production rate upon an inverse heat shock. The response is simulated using the parameter values in Table~\ref{tbl:parameter} except for $F(T)$ which is reversed (suddenly decreased at $t=0$.).
Both models for $\sigma^{32}$-DnaK and HrcA-GroE systems showed very similar responses, which fit well 
with experimental results; the chaperon production rate shows a rapid transient decrease and recovers slower compared with a direct heat shock response.
The fact that the model works without specific tuning of parameters to the inverse heat shock supports that our simple models hit the essential points of the actual reaction mechanisms.

Why the recovery to the new steady state is slower in the inverse heat shock in both systems? The explanation naturally emerges from our model: as chaperons are stable proteins,
the only way to recover to a new steady state upon decrease in unfolded proteins is by dilution due to cell division. Thus this slow time scale for the recovery is given by $1/\gamma_s$ or the time scale of cell division in both systems. In case of direct heat shock the time to recover to new steady state (right after the peak, once there are enough chaperons produced) is governed by the turnover time for active TFs, $\gamma_c$, which is much faster than the rate governed by cell doubling time $\gamma_s$.

\subsection*{
The difference in negative feedback architectures requires different constraints on
TF-chaperon binding affinities
}

The reaction mechanisms of the two systems resemble each other in that there exists a negative feedback loop between a transcription factor and a chaperon.  However, loops are organized such that TF is an activator in one and is an inhibitor in another. In the following we will demonstrate how this difference leads to the distinctly different binding affinities of TF to chaperons.

The constrains on binding affinities can be understood when we look at how TF and chaperons are related in steady state.
From eqs.~(\ref{s1}), (\ref{s4}) and (\ref{s6}) we obtain the expression for free  $\sigma^{32}$ to be
\begin{equation}
\label{eq:ss_sf}
[\sigma^f] = \frac{\eta K_j}{\gamma_s K_j + (\gamma_c +\gamma_s) [D^f]} = \frac{\eta/\gamma_s}{1+\frac{[D^f]}{\gamma_s K_j / (\gamma_c +\gamma_s)}}.
\end{equation}
This is a decreasing function of the free chaperon $[D^f]$, and $[\sigma^f]$ approaches
a constant value $\eta/\gamma_s$ when $[D^f] \ll \gamma_s K_j /(\gamma_c +\gamma_s)$.
Here, $\gamma_s K_j /(\gamma_c +\gamma_s) \sim 3 $ as we fix to $\gamma_s =0.03$ and $K_j = 100$ based on experimental observations.

Similarly for the HrcA, from eq.~(\ref{h1}), (\ref{h4}) and (\ref{h6}), we find
\begin{equation}
\label{eq:ss_ha}
[H^a] = \frac{\beta_h [H^t] [G^f]}{\gamma_c K_l + (\gamma_c +\beta_h) [G^f]} = \frac{\ \ \beta_h [H^t] [G^f] / \gamma_c K_l \ \ }{1+\frac{[G^f]}{\gamma_c K_l / (\gamma_c +\beta_h)}}.
\end{equation}
 This is an increasing function of the free chaperon $[G^f]$, and $[H^a]$ approaches
a constant value $\beta_h [H^t]/ (\gamma_c +\beta_h)$ when $[G^f] \gg \gamma_c K_l / (\gamma_c +\beta_h) $.
Note that both $K_l$ and $\beta_h$ are unknown parameters for the HrcA system.

\begin{figure}
\begin{center}
\centerline{\epsfig{file=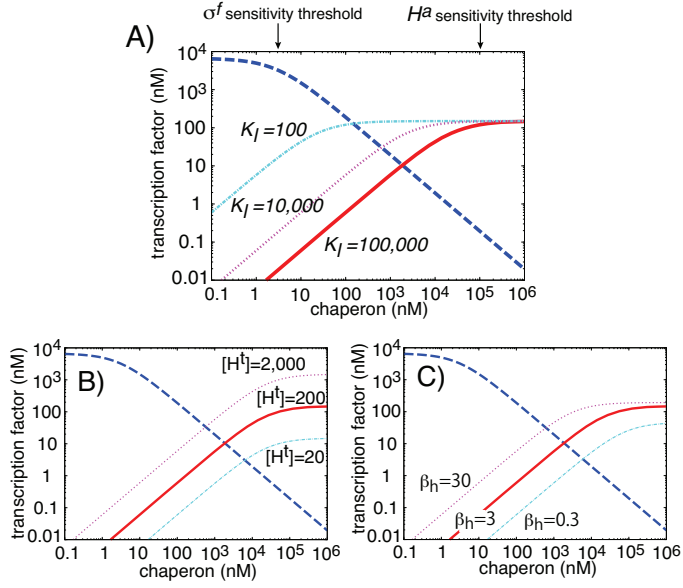,angle=0,width=9cm}}
\caption{Relation between free chaperon and transcription factor in a steady state given by eq.~(\ref{eq:ss_sf}) and eq.~(\ref{eq:ss_ha}). $[\sigma ^f]$ is with the parameters in Table~\ref{tbl:parameter} (bold broken line), while for $[H^a]$ some lines changing a parameter ($K_l$ in A, $[H^t]$ in B, and $\beta_h$ in C) is shown and $\beta_h =3, [H^t]=200$, and $K_l=100,000$ are used if not otherwise specified}
  \label{fig:function}
\end{center}
\end{figure}

These expressions (with parameters from Table~\ref{tbl:parameter}) are plotted in figure~\ref{fig:function}A as functions of free chaperons. 
As chaperons activate HrcA and inhibit $\sigma^{32}$, HrcA is increasing and $\sigma^{32}$ is decreasing with increasing amounts of  free chaperons. 
Each of the curves has two characteristic regimes: a) {\it Insensitive regime}  where  TF are insensitive to changes in chaperone concentration. This corresponds to a nearly flat region in the plot, where concentration of TF does not depend or depends very weakly on the chaperone concentration (chaperones$<3$ for $\sigma^{32}$, and chaperones$>10^5$ for HrcA) and b) {\it Sensitive regime} where change in chaperone concentration results in a change in TF concentration (chaperones$>3$ for $\sigma^{32}$, and chaperones$<10^5$ for HrcA). 

For the system to be responsive and adjust the production rate of chaperons (controlled by amounts of active TFs) in response to changes in unfolded proteins (reflected by the amounts of free chaperons), it is essential for the system to function within sensitive regimes.
The peculiar feature of the two systems is that the insensitive regimes lie in the opposite ends of chaperone concentrations;  $\sigma^{32}$
system is sensitive as long as the free chaperon $[D^f]$ is {\it above} the threshold concentration, 
while HrcA system can work as long as the free chaperon $[G^f]$ is {\it below} the threshold concentration. 
At the same time, the maximal concentration of free chaperon is limited by the amount of total proteins ($\sim 20,000$), and the minimum represents the case when all chaperones are bound to unfolded proteins.
As in steady state the amounts of free chaperones vary between $5,000$-$10,000$, it is crucial that the sensitive regime spans this range.
As insensitive regime for $\sigma^{32}$ lies in the range of small concentrations (threshold $\gamma_s K_j /(\gamma_c +\gamma_s) \sim 3$ with experimentally evaluated parameters), it will always be in the sensitive range.
On the contrary, in HrcA system, the threshold must be larger than the typical steady state concentrations of free chaperons, i.e. $\gamma_c K_l /(\gamma_c +\beta_h) > 10,000$. This impose the condition on binding affinity to be large enough, $K_l > 10,000$, because
the threshold value mainly depends on $K_l$ and tuning other parameters such as $\beta_h$ or $[H^t]$ does not affect threshold values much (figure~\ref{fig:function}B and C).

It now becomes clear why we could not obtain the responses in HrcA with $K_l \sim K_j \sim 100$, as this tight binding in HrcA system would decrease the sensitivity regime to be below $100$ nM or below typical steady state concentrations of free chaperons.
On another hand, for a similar reason, we can not use weak binding affinity for $\sigma^{32}$, i.e. $K_j =100,000$, as it will shrink the sensitivity region to be above $3,000$ nM.

Alternatively, if in reality $K_l \sim 100$ nM this would imply that the sensitivity regime is very narrow, which means that the steady states of free chaperones have to vary between $1$ and $100$ nM.
In principle this could be the case, however, this would imply that the system is not robust to sudden increases in unfolded proteins.

This difference in the chaperone-TF dissociation constant between the two systems is very critical.
While both systems have negative feedback as a core regulatory mechanism, our model predicts that the details of how each of the feedbacks is realized result in very different dissociation constants.
One can test this prediction experimentally by varying binding
affinity in HrcA system or sensitivity domains and investigate how
this affects response dynamics.

\subsection*{Parameter Robustness}

\begin{figure}
\begin{center}
\centerline{\epsfig{file=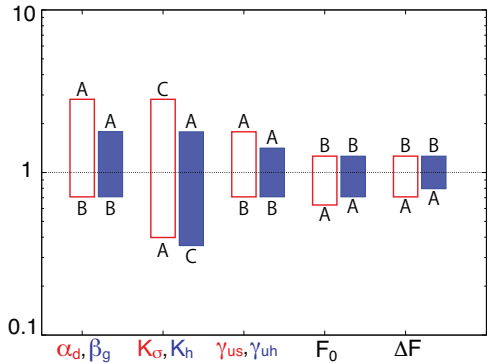,angle=0,width=6.5cm}}
\caption{Parameter robustness measured as fold change over the reference parameter set (shown in Table~\ref{tbl:parameter}) that is able to reproduce a reasonable heat shock response (defined by conditions (i)-(iv)). E.g. the value of $1$ on y axis correspond to the values shown in Table~\ref{tbl:parameter}. Red marks results for $\sigma^{32}$-DnaK system and blue for HrcA-GroE system. $F_0$ means the initial value and $\Delta F$ means fold change in $F(T)$. Capitals at the bottom (top) of each of the bars show what conditions (A,B,C) where broken by further decrease(increase) in parameters. (A) - peak amplitude, (B) - time, and (C) - recovering conditions in chaperon production rate (see text for detail). }
  \label{fig:parameter}
\end{center}
\end{figure}

The results presented so far were based on a single set of parameters, chosen to reproduce experimental data.
Some of them are fixed to a known experimental values as already mentioned in the model section, 
while the rest of the parameters are fitted to reproduce heat shock dynamics; 
there are $5$  fitting parameters for $\sigma^{32}$-DnaK system and $7$ for HrcA-GroE (see Table~\ref{tbl:parameter}).
To understand how constrained is our parameter choice we studied robustness of parameters.
Figure~\ref{fig:parameter} shows how much a given parameter can be changed (the maximal fold of change from the values shown in Table~\ref{tbl:parameter}) with preserving proper heat shock response.

We choose a proper response to the one characterized by  (A) a peak with more than $2$ and less than $10$ fold change (normalized to the pre-stimulus level) (B) occurring within $10$ unit times and (C) recovering to a new steady state that is less than $2$ fold of the pre-stimulus one in chaperon production rate.
 Interestingly,  
most parameters can only be changed at most a few fold for both systems.

One of the main reasons why system is rather sensitive to parameter choice is due to the importance of the stoichiometry between chaperons and unfolded proteins: if either chaperons or unfolded proteins are in excess there will be no peak.  Excess of chaperons (e.g. due to high chaperone production rate $\alpha_d, \beta_g$) will absorb a sudden increase in unfolded proteins, so that
there will be no increase in chaperone production and thus no peak. On the other hand excess of unfolded proteins (e.g. low $\alpha_d, \beta_g$) will result in a state where chaperone production is maximally activated already before the shock. Thus a further increase in unfolded proteins will not lead to the increase in production rate of chaperons.

\section*{Discussion} 

We have quantitatively investigated similarities and differences in two heat shock systems:
one characteristic to gram negative bacteria (e.g. \textit{E. coli}) and another to gram positive (e.g. \textit{L. lactis}). 
Remarkably, although the two are very different at the level of promoter regulation, a striking similarity appears at the level of regulatory networks. Both are governed by chaperon-mediated negative feedback loops and in both cases chaperon sequestration is employed as a stress sensing mechanism. Furthermore the similarity continues at the level of the response dynamics -- both systems have characteristic rapid transient responses.

There are three core features characteristic to both systems, that are necessary to generate a rapid transient response observed in both systems upon direct heat shock ( a sudden increase in unfolded proteins).

\begin{itemize}
\item[a)] The initial rapid increase is governed by chaperone independent rates,
which are the $\sigma^{32}$ synthesis rate or the  rate of HrcA conversion into inactive form. The initial slope of increase in chaperons is governed by respectively  $\eta$ and $\gamma_c$.
\item[b)] The rapid recovery to steady state is governed by chaperone mediated processes (degradation of $\sigma^{32}$ or activation of HrcA)

\item[c)] The peak is the result of a two rather different time scales involved:  a rapid dynamics of TF (determining rapid increase and decrease) and a slow chaperone synthesis, determining the time of the peak in transient response, i.e. time when there is enough chaperones to deal with increased amounts of unfolded proteins.  
\end{itemize}

Furthermore, different realizations of negative feedbacks -- one through an transcriptional activator another through transcriptional inhibitor -- impose distinct constrains on chaperone-TF binding affinities. Our analyses predicts that whereas the tighter TF-chaperone binding increases dynamic range for $\sigma^{32}$ system, it would work in opposite direction and decrease the dynamic range for HrcA system. 
In other words,  chaperone-TF binding affinity imposes a lower limit on the amounts of free chaperones
for $\sigma^{32}$ system, where it becomes the upper limit for HrcA system.
With the experimentally determined binding affinity for $\sigma^{32}$ system, the lower limit for free chaperons is $1$ nM (see figure~\ref{fig:function}), i.e. one free chaperone per cell, which is low enough to account for possible variations in chaperone levels. (The upper bound in this case is determined by the amounts of TF to be not less than $1-2$ protein per cell ($1-2$ nM) so that each cell feels change in  TF, thus setting upper limit to about $10^4$, see figure~\ref{fig:function}.)

We predict that the chaperone-HrcA binding should match the upper limit of the desired amount of free chaperons and when measured can thus serve as indirect indication of the amounts of free chaperones in the cell.

\section*{Acknowledgement} 

We thank  Kim Sneppen and Steen Pedersen for useful discussions. The research was supported by the Danish National Research Foundation.  MI is supported by JSPS Research Fellowships for Young Scientists and AT us supported by Steno fellowship.

\bibliographystyle{unsrt}
\bibliography{HeatShockAla}
\end{document}